\documentclass[%
preprint,
superscriptaddress,
longbibliography,
 amsmath,amssymb,
 aps,
]{revtex4-1}

\usepackage{graphicx}
\usepackage{dcolumn}
\usepackage{bm}
\usepackage{tabularx}
\usepackage{hyperref}
\newcommand{\abs}[1]{\left\vert#1\right\vert}       
\newcommand{\var}[1]{\left(#1\right)}               
\newcommand{\com}[1]{\left[#1\right]}               
\newcommand{\mr}{\mathrm}                           
\newcommand{\mb}{\mathbf}                           
\newcommand{\up}{\uparrow}
\newcommand{\dw}{\downarrow}

\begin{document}

\title{Control of superconductivity with a single ferromagnetic layer in niobium/erbium bilayers}

\author{N.~Satchell}
\affiliation{School of Physics and Astronomy, University of Leeds, Leeds, LS2 9JT, United Kingdom}
\affiliation{ISIS Neutron and Muon Source, STFC Rutherford Appleton Laboratory, Chilton, Didcot, Oxon, OX11 0QX, United Kingdom}

\author{J.~D.~S.~Witt}
\affiliation{School of Physics and Astronomy, University of Leeds, Leeds, LS2 9JT, United Kingdom}

\author{M.~G.~Flokstra}
\affiliation{School of Physics and Astronomy, SUPA, University of St Andrews, St Andrews KY16 9SS, United Kingdom}

\author{S.~L.~Lee}
\affiliation{School of Physics and Astronomy, SUPA, University of St Andrews, St Andrews KY16 9SS, United Kingdom}

\author{J.~F.~K.~Cooper}
\affiliation{ISIS Neutron and Muon Source, STFC Rutherford Appleton Laboratory, Chilton, Didcot, Oxon, OX11 0QX, United Kingdom}

\author{C.~J.~Kinane}
\affiliation{ISIS Neutron and Muon Source, STFC Rutherford Appleton Laboratory, Chilton, Didcot, Oxon, OX11 0QX, United Kingdom}

\author{S.~Langridge}
\affiliation{ISIS Neutron and Muon Source, STFC Rutherford Appleton Laboratory, Chilton, Didcot, Oxon, OX11 0QX, United Kingdom}

\author{G.~Burnell}
\email{g.burnell@leeds.ac.uk}
\affiliation{School of Physics and Astronomy, University of Leeds, Leeds, LS2 9JT, United Kingdom}

\date{\today}

\begin{abstract}

Superconducting spintronics in hybrid superconductor--ferromagnet (S--F) heterostructures provides an exciting potential new class of device. The prototypical super-spintronic device is the superconducting spin-valve, where the critical temperature, $T_c$, of the S-layer can be controlled by the relative orientation of two (or more) F-layers. Here, we show that such control is also possible in a simple S/F bilayer. Using field history to set the remanent magnetic state of a thin Er layer, we demonstrate for a Nb/Er bilayer a high level of control of both $T_c$ and the shape of the resistive transition, R(T), to zero resistance. We are able to model the origin of the remanent magnetization, treating it as an increase in the effective exchange field of the ferromagnet and link this, using conventional S--F theory, to the suppression of $T_c$. We observe stepped features in the R(T) which we argue is due to a fundamental interaction of superconductivity with inhomogeneous ferromagnetism, a phenomena currently lacking theoretical description. 

\end{abstract}


\maketitle

\section{Introduction}
While traditionally considered competing phenomena, when artificially juxtaposed, there is a wealth of physics at the interface between superconductors (S) and ferromagnets (F). Taking advantage of the competition between order parameters has lead to advances in the emerging field of super-spintronics \cite{linder_superconducting_2015}. By placing an inhomogeneous magnetic texture at the S--F interface, it is possible to create the so-called long ranged triplet component (LRTC) or finite spin Cooper pair. Unlike the singlet Cooper pair, the LRTC is not dephased by the exchange field and can therefore penetrate further into a proximitised F-layer. This opens the exciting possibility of performing spintronic logic operations on a dissipationless spin current \cite{eschrig_spin-polarized_2011}. Additionally, several breakthroughs in complex S--F heterostructures show promise as potential cryogenic memory elements. In such a scheme, information could be stored by the state of the system (superconducting or normal) \cite{oh_superconductive_1997, PhysRevB.71.180503} or the ground-state phase difference between two S-layers in an S/F/S Josephson junction \cite{bell_controllable_2004, goldobin_memory_2013, baek_hybrid_2014, niedzielski_use_2014, gingrich_controllable_2016}.

The prototypical super-spintronic device is the superconducting spin valve. In this device, control of the magnetic state of the two F-layers in an S/F/F or F/S/F heterostructure can be used to tune the generation of the LRTC \cite{leksin_evidence_2012, jara2014angular, wang_giant_2014, flokstra_controlled_2015, flokstra2016remotely, PhysRevB.93.100502}. The generation of the LRTC opens an additional conduction channel for Cooper pairs, resulting in the lowering of $T_c$ \cite{fominov2010superconducting}. In our previous work we found this suppression of $T_c$ to be of the order 10--20~mK in a 3$d$ ferromagnet/niobium device \cite{flokstra_controlled_2015}, although other works have increased this effect; to 130--140~mK by carefully engineering both the superconducting layer and S--F interface \cite{wang_giant_2014, PhysRevB.93.100502}, and to over 1~K by both introducing a half-metal as the bottom F-layer and changing the applied field orientation from an in-plane rotation to an in-plane to out-of-plane rotation \cite{singh2015colossal, PhysRevB.94.054503}. The manipulation of the F-layers in the superconducting spin valve requires careful engineering of the heterostructure and the rotation of the sample in an applied magnetic field. Under an in-plane field rotation it is possible to introduce experimental artefacts due to: vortex flow (if too high a current is applied, induced voltage from vortex flow will dominate the transport signal); non-uniformity of field (if the sample is not aligned correctly the out-of-plane field component will vary under rotation - modifying $T_c$); and temperature (a temperature gradient or local source of heating inside a cryostat is an important consideration when the sample is moving during measurement). Any of these can introduce a signal with the same periodicity as the signature of LRTC generation. A recent theoretical work considered that there exists a ``half-select" problem in the multilayer spin valve approach, which may be negated in a simplified device \cite{pugach2017superconducting}. 

In this work we describe a simplified S--F hybrid system, where the superconductivity can be controlled by a single adjacent F-layer. The system only requires the ability to apply an external field in one direction (without the need for sample rotation) and we perform all our measurements in zero applied field, two distinct advantages over the superconducting spin valve. This is achieved by coupling a superconducting Nb layer to rare-earth ferromagnetic Er, which has a large number of metastable magnetic phases accessible with temperature or applied field. Previous work on holmium and dysprosium demonstrate the important role rare-earth ferromagnets will play in the implementation of superconducting spintronics in Josephson type devices \cite{PhysRevLett.96.157002, robinson2010controlled} and devices based on the control of $T_c$  \cite{gu_magnetic_2014, PhysRevLett.115.067201}. For example, Gu \emph{et al.} demonstrated that an antiferromagnetic to ferromagnetic transition in Ho resulted in modification of the $T_c$ of an adjacent Nb layer of over 100~mK, however the exact mechanism involved in the $T_c$ shift was not established \cite{gu_magnetic_2014}.  This work was later expanded by producing trilayer samples of Ho/Nb/Ho and Dy/Nb/Dy in which a spin valve like effect of 400~mK was discovered \cite{PhysRevLett.115.067201}. These works established the ability to control $T_c$ with the ferromagnetic texture in rare-earth ferromagnets, however lacked the theoretical description and the additional modification to the shape of the R(T) transition reported in this manuscript.  

Er is a trivalent rare-earth metal (Z=68), with highly localised 4-f electrons and a hexagonal close packed (hcp) crystal structure. 
Competition between the RKKY indirect exchange interaction and the crystalline anisotropy, creates a rich magnetic phase diagram making this material ideal for the exploration of S--F proximity effects \cite{habenschuss_neutron_1974,gibbs_magnetic_1986,lin_magnetic_1992,cowley_magnetic_1992,Jehan_collapsing_1994, watson_magnetic_1995,watson_b-axis_1996,frazer_magnetic_1999}. 
Below the high-temperature paramagnetic phase ($\approx85$~K), Er first gains a sinusoidal, $c$-axis modulated (CAM) anti-ferromagnetic phase. As the temperature is lowered, the magnetic wave vector of the CAM expands until $\approx52$~K. Below this temperature Er enters an `intermediate' phase where the in-plane moments begin to order creating what has been referred to by Cowley \emph{et al.} as an anti-ferromagnetic ``wobbling cycloid'' \cite{cowley_magnetic_1992}. The magnetic cycloid repeat distance increases with decreasing temperature, through a number of stable commensurate phases, to 8 atomic layers. These states exhibit a ferrimagnetic moment. Finally, below 18~K a conical $c$-axis ferromagnetic phase is formed. We have been able to confirm that many of the magnetic states of bulk Er are reproducible in sputter deposited epitaxial thin films, and that these magnetic states can be controlled with either temperature or applied magnetic field \cite{wittscientificreports, satchellpnr}.

\section{Methods}
The films were deposited using DC sputtering in a system with substrate heaters mounted above each sample slot. At the highest temperature, the base pressure of the system is $\approx 10^{-7}$~mbar. This pressure improves as the system temperature is lowered. The samples were grown on 0.65~mm thick $c$-plane Al$_2$O$_3$ substrates. The Nb was deposited at a nominal temperature of 700$^{\circ}$C, after which the system was cooled to 500$^{\circ}$C, a final thin Nb interface layer was deposited at this temperature, followed directly by the Er and then a 5~nm-thick Lu capping layer.  Growth was performed at a typical Ar flow of 55~sccm resulting in Ar pressure of 2-3~$\mu$bar, at a substrate--sample distance of 75~mm, and at a typical growth rate of 0.1~nm~s$^{-1}$. Growth rates were calibrated by fitting to Keissig fringes obtained on single layer samples by X-ray reflectometry. The Nb was grown first as it has been shown to be an effective buffer layer for the growth of rare-earth metals and stops the Er layer reacting with oxygen in the substrate \cite{kwo_growth_1986}. The Nb/Er interface is known to be sharp due to the lack of alloying and intermixing between Nb and Er \cite{moffatt_handbook_1981}. The Er grows epitaxially on the most densely packed Nb (110) plane, in the Nishiyama-Wasserman orientation. The in-plane axis of hcp Er [10$\bar{1}$0] is aligned with bcc Nb [$\bar{1}$10] with 3:4 supercell commensuration in their nearest-neighbour distances along these axes \cite{kwo_growth_1986}. Lu was chosen for the capping layer as it lattice matches well with Er (preventing additional strain being introduced), and unlike some traditional capping metals, such as Au, it can be deposited as a continuous layer at high temperatures \cite{wittscientificreports}.

Magnetization loops and remanent magnetization, $M_r$, were measured using a 6~T Quantum Designs SQUID-VSM magnetometer at 10~K. Electrical transport measurements were performed on sheet films using a conventional four point probe measurement configuration and employing two continuous flow $^4$He cryostats, with maximum fields of 3~T and 8~T. The field histories were only applied when the sample was in the normal state (to prevent flux trapping). The resistance as a function of temperature (R(T)) of the sample, from which $T_c$ is obtained, was always measured at zero applied field. Temperature sweeps, both cooling and warming, were recorded to check for temperature hysteresis in the measurements. The temperature hysteresis (observable in FIG.~\ref{RvT}) does not account for the observed $T_c$ shift in FIG.~\ref{dTc}. 

\section{Results}

\subsection{Magnetic Characterization}

\begin{figure} 
\includegraphics[width=0.9\columnwidth]{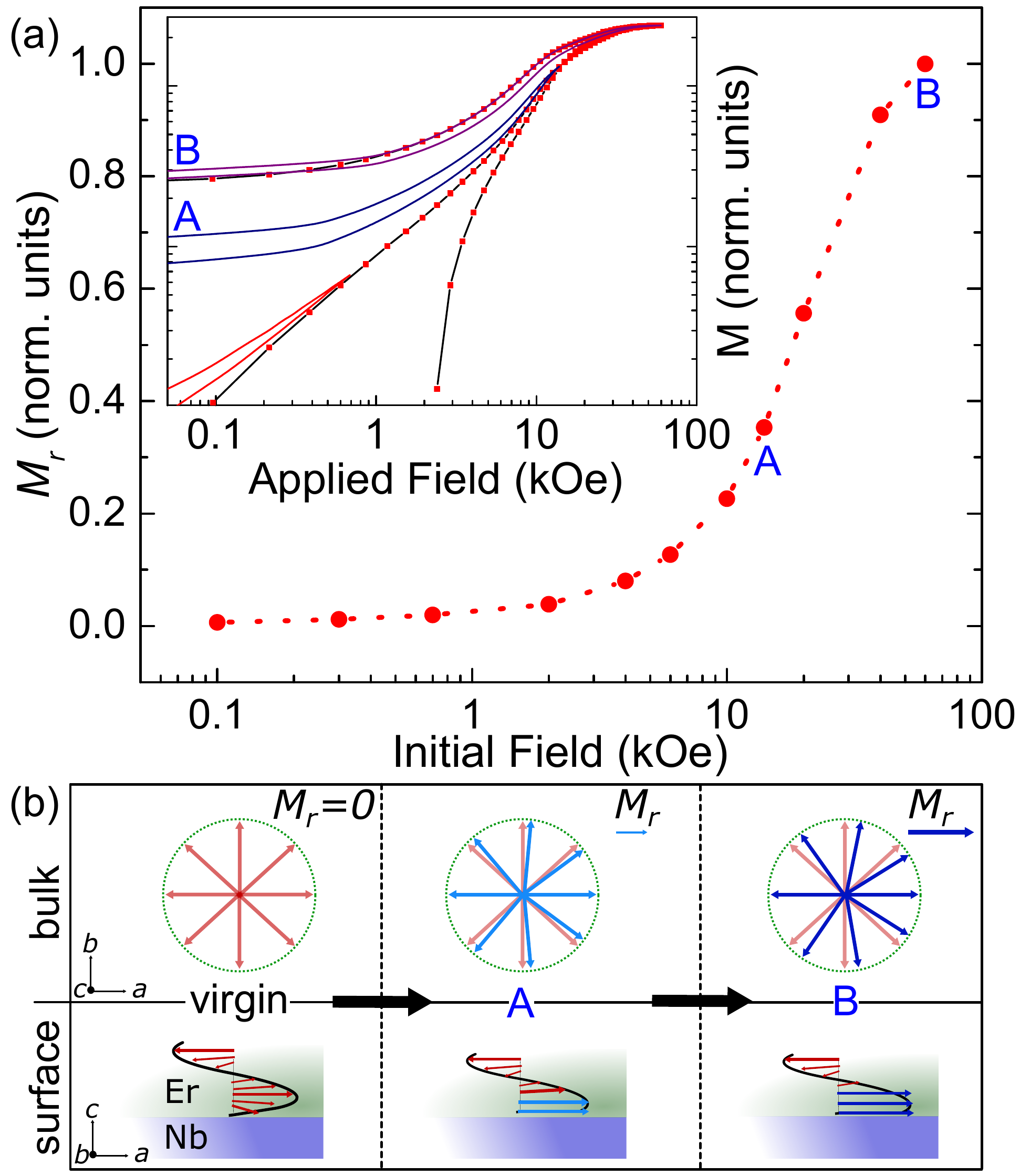} 
\caption{(a) The normalised remanent magnetization, $M_r$, as a function of initial field. Inset: Positive quadrant of the magnetic hysteresis loop (squares) and exemplar minor loops (solid lines) of the bilayer at 10~K. The data are displayed on a log-log plot for increased clarity. (b) The two possible mechanisms to account for $M_r$. The bulk mechanism shows a canting of the spiral into the direction of the applied field. The surface mechanism shows an interface effect, where only the surface moments remain aligned in the direction of the applied field. \label{loops}}
\end{figure}

The magnetization versus field data, along with minor loops, for the Nb(20~nm)/Er(25~nm) bilayer sample at 10~K are shown as the inset in FIG.~\ref{loops} (a). The red squares show the initial magnetization and full magnetic hysteresis behaviour for applied magnetic fields up to 60~kOe. The solid lines are a series of minor loops, from which information about the $M_r$ of the sample can be obtained. The $M_r$ as a function of initial field data are collated in FIG.~\ref{loops} (a).

It is evident from FIG.~\ref{loops} (a) that for low initial fields, there is little change to the remanent state of the Er. This indicates that, in this range, the stabilisation of the spiral magnetic structure in the Er---due the RKKY interaction---is robust against perturbation by the externally applied magnetic fields. The large increase in $M_r$ for initial fields of about 10~kOe is evidence that, for initial field values greater than this, the Er does not re-enter the same magnetic phase upon relaxation of the field. This is consistent with previous characterization work which shows that at approximately 25~kOe there is a phase transition, for an in-plane applied field, into a distorted spiral phase, known as the `fan' or `canted-fan' state \cite{wittscientificreports, frazer_magnetic_1999}. Possible origins of the increased remanence are shown schematically in FIG.~\ref{loops} (b) and discussed further in section IV A.

\subsection{Electrical Transport}

\begin{figure}
\includegraphics[width=0.9\columnwidth]{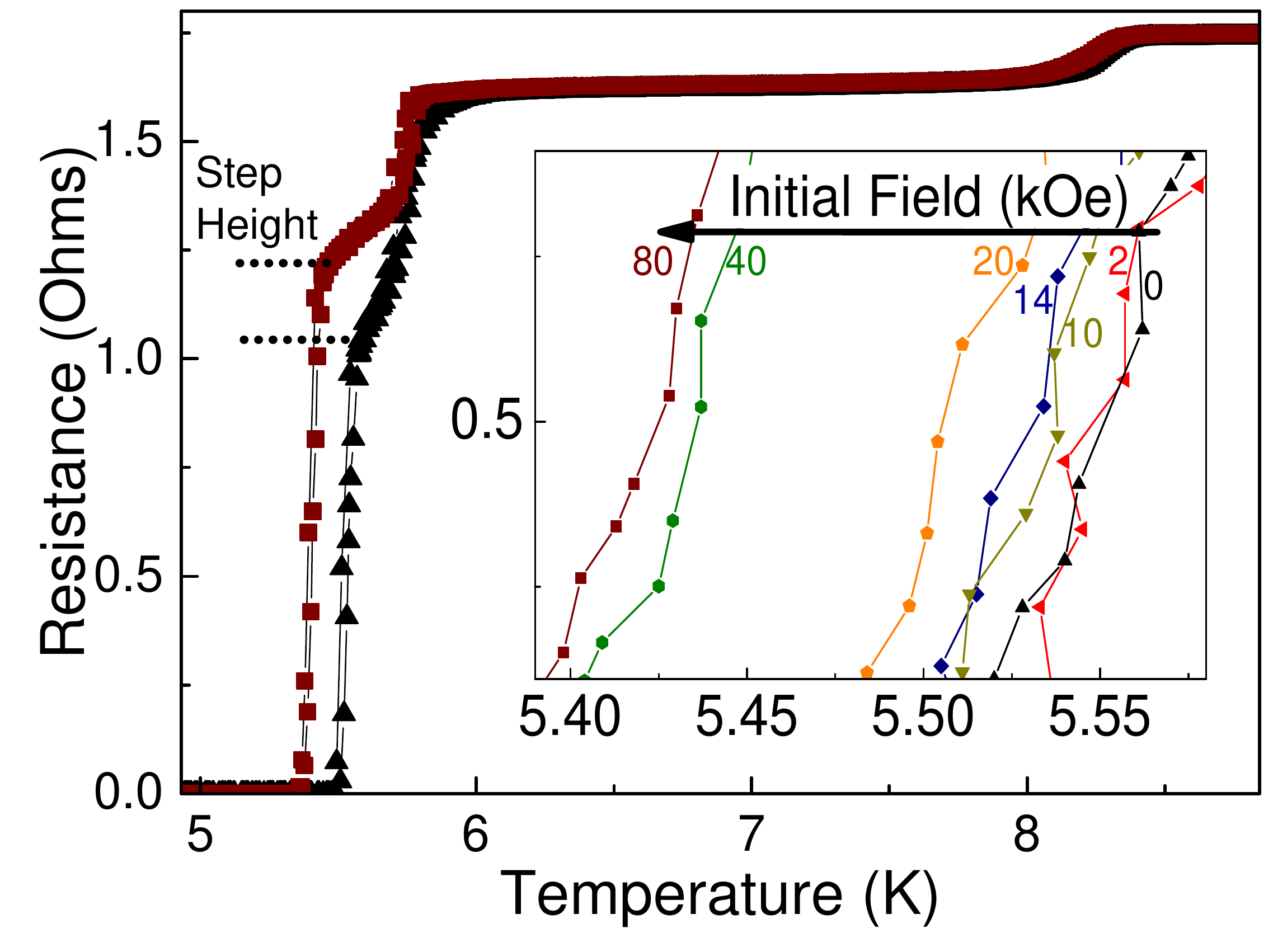}
\caption{Resistance as a function of temperature (R(T)) for the virgin magnetic state (triangles) and post 80~kOe field saturation (squares). Both warming and cooling data are included and the changing step heights marked. Inset: Exemplar cooling curves demonstrating the evolution of R(T) with initial field. \label{RvT}}
\end{figure}

\begin{figure}
\includegraphics[width=0.9\columnwidth]{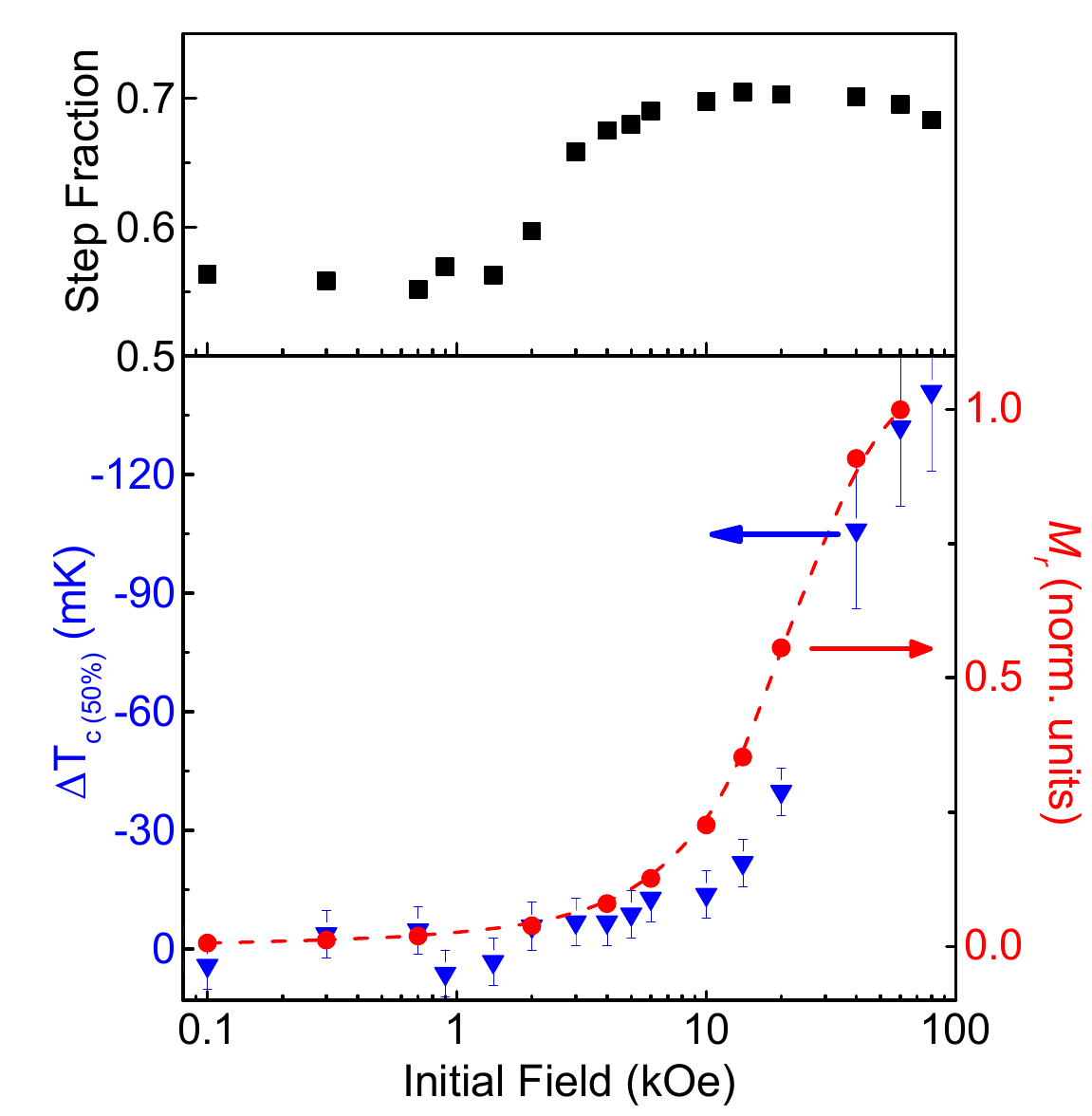}
\caption{Top panel: The fractional step height change with initial field. Bottom panel: The normalised remanent magnetization, $M_r$, (circles - reproduced from FIG~\ref{loops}) and the shift in superconducting critical temperature from the as cooled state, $\Delta T_c$, (triangles) as a function of initial applied magnetic field. \label{dTc}}
\end{figure}

FIG.~\ref{RvT} shows the resistance as a function of temperature  for the Nb(20~nm)/Er(25~nm) bilayer sample, always measured in zero applied field. The data show the onset of superconductivity as the temperature is decreased for the virgin state (triangles) and after the application of an 80~kOe applied magnetic field (squares). In the inset of FIG.~\ref{RvT} the evolution of R(T) as a function of the initial applied magnetic field can be seen. Resistance as a function of temperature was always measured in zero applied magnetic field and $T_c$ was defined as $50 \%$ of the normal state resistance. $\Delta T_c$ is calculated as the difference between $T_c$ of the virgin state (triangles in FIG.~\ref{RvT}) and $T_c$ after the application and removal of a magnetic field. The $\Delta T_c$ data for all of the initial applied magnetic field values are collated in FIG.~\ref{dTc}.

In FIG.~\ref{dTc} it is immediately clear that there is a strong link between the $M_r$ of the Er and the $T_c$ of the superconductor. This correlation, between the properties of Er and Nb, show that both the $T_c$ of the Nb and the magnetic state of the Er are strongly dependent upon the field history of the sample. It also shows that there is a strong coupling between the superconducting and magnetic layers. The largest change to $\Delta T_c$ comes between 20--30~kOe, which, as mentioned above, is also the field value where the Er state changes magnetic phase. After the application of the largest field possible in our system, 80~kOe, the $T_c$ of the Nb was suppressed by approximately 140~mK, which is the largest value reported for such a system. The metastable magnetic state obtained by applying and removing and initial field, is robust against temperature changes in the measured range 5-10~K. The system can be effectively reset to the virgin state by warming through the Curie temperature.

One additional point to note is the step-like features that are present in FIG.~\ref{RvT}, and the fact that these steps also change with field history. The height of the stepped feature in the transition is marked on FIG.~\ref{RvT}. A step fraction is defined as the height of the step relative to the normal state resistance (at 10~K). The collated step fraction is plotted in the upper panel of FIG.~\ref{dTc} and is discussed further in section~\ref{discussion}.


\section{Modelling}





\subsection{Modelling of $M_r$: An Effective $E_{ex}$}

Having established that the suppression observed in $T_c$ is linked to the increased $M_r$, we now consider the local magnetic state of the Er film and propose two physical interpretations for the origin of $M_r$ in Er. The first being a `bulk' modification of the spiral and the second a localised spin alignment at the Er surface. 

As we have shown in our previous work, even in the thin film, Er has a highly complicated phase diagram \cite{wittscientificreports, satchellpnr}. Through a combination of temperature and field, the Er can be placed in a number of metastable magnetic states. For an in-plane field, between 20 and 30~kOe Er undergoes a transition from the conical to a `fan' magnetic state, canted into the direction of applied field. Subsequent removal of the applied field causes the Er to re-enter the conical state, however we argue the cone has now been modified and is canting in the direction of the external applied field, increasing $M_r$. This canted conical state is shown schematically as the top mechanism in FIG.~\ref{loops} (b).

It is well known that finite-size effects play an important part for thin film rare-earths \cite{bohr1989diffraction}. The long-range nature of the RKKY interaction (up to 6th nearest neighbour) means that the reduced atomic coordination at the surfaces makes the spiral ends less robust against external perturbation, which clearly becomes more of an influencing factor for thinner films with a lower volume to surface area ratio \cite{satchellpnr}. It is, therefore, possible that, under the influence of an externally applied magnetic field, the spiral unwinds more readily at the surfaces and, being unable to overcome the energy barrier to reform the spiral, remains locally in the direction of the applied field. This is shown schematically as the bottom mechanism in FIG.~\ref{loops} (b). We calculate from the known thickness, saturation magnetization and expected moment per atom that 0.65~nm (or just over two atoms) remaining aligned at the interface would account for the observed $M_r$ (see SI FIG~1).




The `bulk' canted magnetic phase which leads to a net magnetization could be described by an effective exchange field if the coherence length inside the Er is (much) longer than the magnetic repeat unit of the helix, which is about 4~nm. On the other hand, the contribution of the surface moments to the total proximity effect is only considerable if the coherence length is short, comparable to the 0.65~nm effective Er thickness corresponding to aligned surface moments. The two mechanisms thus correspond to very different length scales of the coherence length inside the Er layer.


Resistance measurements on Er/Nb bilayers with various Er layer thicknesses suggest an approximate coherence length of 10~nm (see SI FIG~2) and in a related Ho system the coherence length was estimated to be 30~nm \cite{PhysRevLett.115.067201}. This distance is far greater than 0.65~nm and is long enough to allow the Cooper pair to experience multiple helicies. It is, therefore, most likely that the Er undergoes the `bulk' transition to a new canted magnetic phase, which is retained at zero field, and that this is the origin of the suppression in $T_c$.

\subsection{Modelling of $\Delta T_c$ vs.~$M_r$}  

\begin{figure}
\includegraphics[width=1\columnwidth]{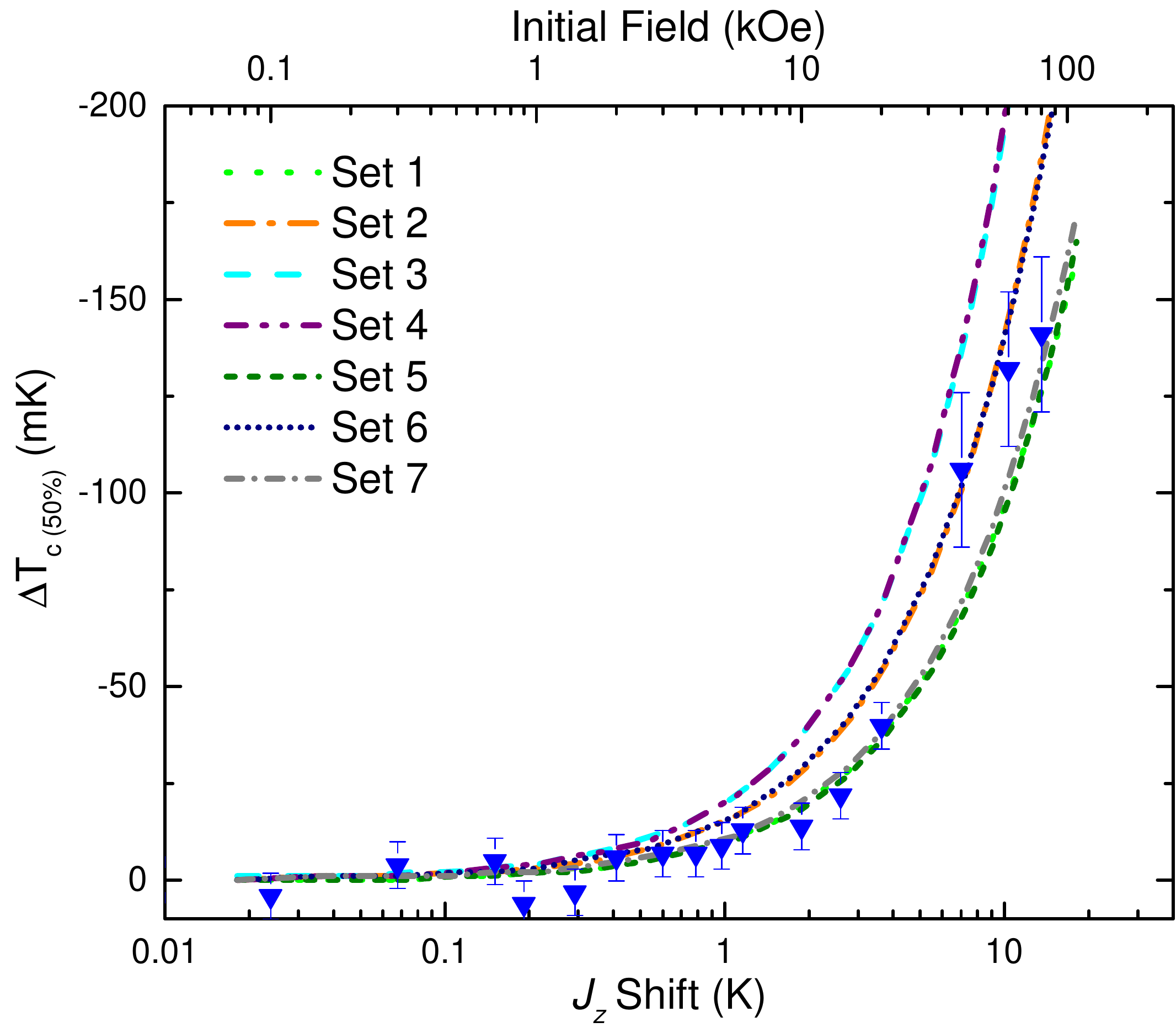}
\caption{The calculated suppression of $T_c$ with an effective shift to the exchange field ($J_z$). Also plotted are the $\Delta T_c$ data from FIG. \ref{dTc}. The parameter `sets' used in the calculation are defined in the text and listed in the SI Table 1. \label{TcdJ}}
\end{figure}


To investigate the effect of an increased bulk remanent magnetisation, we model it as an effective exchange field ($J_z$) in the F layer, which we can then link directly to the suppression in $T_c$.

Using the quasiclassical theory for superconductivity in the dirty limit (electronic mean free path much shorter than the phase coherence length), we calculate the critical temperature of a Er/Nb bilayer as function of an effective exchange field inside the Er. We take the x-axis normal to the metallic layers and assume translational invariance in the y,z plane. The Usadel equation for $s$-wave superconductivity then takes the form $i\hbar D\partial_x\var{\check{g}\partial_x\check{g}} = \com{\check{H},\check{g}}$ with $\check{g}$ the $4\times 4$ matrix Green function in the Nambu spin space, $\hbar$ the reduced Planck constant and $D$ the diffusion constant. For collinear exchange fields the Hamiltonian can be described by $\check{H} = i\hbar\omega_n\var{\tau_3\otimes\sigma_0} + \check{\Delta} - J_z\tau_0\otimes\sigma_3$ (see e.g. \cite{PhysRevLett.95.187003}) with $J_z$ the exchange field directed along the z-axis and $\omega_n$ the Matsubara frequencies defined by $\hbar\omega_n = \pi k_B T\var{2n+1}$ with $k_B$ the Boltzmann constant, $n$ integer, and the maximum allowed frequency given by the Debye frequency. x-y-z is defined such that $J_z$ points along the direction of the net moment of the Er. Furthermore, $\sigma_i$ and $\tau_i$ are the Pauli matrices of respectively the spin space and Nambu space. The matrix Green function and $\check{\Delta}$ have the following non-zero elements:
\begin{equation}
\check{g} = \left(\begin{array}{cccc}
G_{\up\up} & 0 & 0 & F_{\up\dw}\\
0 & G_{\dw\dw} & F_{\dw\up} & 0 \\
0 & \overline{F}_{\up\dw} & \overline{G}_{\up\up} & 0\\
\overline{F}_{\dw\up} & 0 & 0 & \overline{G}_{\dw\dw}\end{array}\right)\mbox{, }\check{\Delta} = \left(\begin{array}{cccc}
0 & 0 & 0 & -\Delta\\
0 & 0 & \Delta & 0\\
0 & -\Delta^* & 0 & 0 \\
\Delta^* & 0 & 0 & 0\end{array}\right)
\end{equation}
where $G$ and $F$ are the quasiclassical normal and anomalous Green functions respectively, both functions of $(x,\omega_n)$, and $\Delta\var{x}$ is the order parameter. The matrix Green function satisfies the normalization condition $\check{g}^2 = \check{1}$ and the order parameter must be solved selfconsistently satisfying the gap equation:
\begin{equation}
i\Delta\var{\mb{R}} = \frac{\pi k_B T}{\ln\var{\frac{T}{T_{c0}}} + \sum_{n}\var{\frac{1}{\abs{2n+1}}}} \sum_{\omega_n}F_{\up\dw}\var{\mb{R},\omega_n} 
\end{equation}
with $T_{c0}$ the bulk critical temperature. We use the interface boundary conditions as formulated by Nazarov\cite{nazarov1999novel} which for the interface between two materials with labels $l,r$ for the layer on the left and right side of the interface respectively can be written as:
$\sigma_l\check{g}_l\partial_x\check{g}_l = \sigma_r\check{g}_r\partial_x\check{g}_r$ and $\sigma_l\check{g}_l\partial_x\check{g}_l = \frac{2}{R_b}\frac{\com{\check{g}_l,\check{g}_r}}{4 + \Gamma\var{\check{g}_l\check{g}_r+\check{g}_r\check{g}_l-2}}$, with $\sigma_i$ the conductivity of layer $i$, $0\leq\Gamma\leq1$ the interface transparency and $R_b$ the interface resistance times the interface area ($\Omega$ m$^2$).\\

The material parameters used for the Nb layer are $\xi_s=\sqrt{\hbar D_s/(2\pi k_BT_\mr{c0})}=7.9$ nm, $T_{c0}=8.4$ K and $\rho_s=15.2$ $\mu\Omega$ cm. Since the value of $J_z$ is unknown we explored various combinations of $\xi_f=\sqrt{\hbar D_f/J_z}$, $J_z$ and $R_b$ chosen such that the $T_c$ of the bilayer corresponds to the experimental value of 5.5 K. For all calculations $\Gamma=1$. For each material combination $T_c$ was calculated as a function of a shift in $J_z$ (a shift of zero corresponding to the $T_c$ of 5.5 K). 

The results of the modelling are presented in FIG. \ref{TcdJ} along with the experimental data. When taken with the thickness dependence, SI FIG.~2, it is parameter set 2 and 5 which show closest agreement to the experimental data, although all parameter sets considered qualitatively reproduce the experimental data. These two parameter sets give the same value of interface resistance, but were considered with different initial values of $J_z$, set 2 corresponding to the lowest temperature ($\approx 22$~K) conical ferromagnetic transition, and set 5 the transition from the antiferromagnetic to ferromagnetic ``wobbling cycloid'' intermediate state ($\approx 55$~K), both of which have been confirmed in our thin films \cite{wittscientificreports}. From either starting point, the analysis shows that the observed 140~mK $T_c$ shift corresponds to a 5-10~K shift in $J_z$ ($7.5-15 \times 10 ^{-20}$~meV).


\section{Discussion}\label{discussion}



Rare earth ferromagnets, such as Er, offer a plethora of magnetic configurations in which a Cooper pair (coherent with a neighbouring proximity coupled superconductor) can experience magnetic disorder. As a conical ferromagnet, Er is a theoretically ideal system in which to generate and study proximity effects induced by an additional LRTC \cite{fritsch_proximity_2014}. In this work it is expected that all R(T) measurements were performed when the Er was in a disordered magnetic state. It is therefore not possible to directly attribute LRTC generation to the observed $\Delta T_c$. In comparison to the superconducting spin valve, which has a clean LRTC on/off mechanism (as magnetic inhomogeneity is carefully engineered from otherwise homogeneous magnetic layers), our proposed origin of $M_r$ cannot provide such a switching mechanism. The canting of the magnetic state into the direction of applied field is unlikely to significantly change the conversion efficiency of singlet Cooper pairs into the LRTC. In the second proposed mechanism, spins at the surface remain aligned with the applied field, and could create a homogeneous interface layer. From the spin valve experimental argument we would expect this to result in a decrease in LRTC generation and therefore an increase in $T_c$. This does not agree with the experimental observation in this work.



Given the size of this $T_c$ effect is generally larger than that reported for spin-valves (and the number of reported cases showing an effect opposite to the spin-valve effect, where the disordered magnetic state results in a \emph{higher} measured $T_c$ \cite{rusanov_enhancement_2004, zhu_altering_2008, zhu_unanticipated_2013, gu_magnetic_2014}), we urge caution for the interpretation of $T_c$ measurements alone as evidence for the presence of the LRTC in S--F systems.


With the modelling we have shown that the reported change in $T_c$ can be described within the conventional S--F proximity theory by considering the increasing remanence of the Er as a shift in the effective exchange field. This increase in exchange field modifies the proximity effect, suppressing the $T_c$ of the bilayer. An effective shift in $J_z$ of 5-10 K accounts for the observed changes in $T_c$. 


In the transition curves, shown in FIG.~\ref{RvT}, three step-like features can be seen. The first, present at $\approx8$~K, appears to be directly related to the $T_c$ of the bare Nb film. This is evidence of local regions of the bilayer film where the Er has no direct influence on the Nb, that is, where the two materials are not coupled by the proximity effect. One possibility for this is at the Er grain boundaries or local regions around the wire-bond contacts, where the force of the contact may have disrupted the Nb/Er interface. This interpretation is supported by the observation that there is no significant field history dependence of this step.


Some common explanations for step-like features in the transition curves can be ruled out for our system. The sputtering technique employed in this work is unlikely to create a significant thickness gradient. To check the uniformity of the films, a 20$\times$20~mm film was diced into several pieces and X-ray reflectivity was performed. A 5\% variation in thickness was observed, this variation is only slightly greater than the error in individually calculated thickness by fitting to Keissig fringes. By comparison, the sample size for transport measurements was only 3$\times$3~mm, where uniformity in film thickness will be very high. Crystallographic inhomogeneity is a further possibility, but again unlikely. We examined a possible current (heating) dependence of the step-like features using currents ranging from 100~nA up to 1~mA, but found no such evidence and current induced local heating can thus be ruled out. There are no known Nb-Er alloys and in our previous reflectivity work we observed no evidence for intermixing at the interface \cite{satchellpnr}, which if possible could have altered the superconducting properties of the Nb. Poor interface transparency can cause anomalous features in resistivity around the superconducting transition, as current paths change to flow preferentially through the superconductor. The formation of an oxide barrier at the interface would cause such effects, but the calculated oxidation time in our vacuum of 15 minutes is far longer than the 20 seconds between the final Nb and Er layer depositions. The steps are never observed in either single layer Nb films (deposited under identical growth conditions), or films of Nb grown in proximity to a homogeneous ferromagnet such as Co (see for example \cite{flokstra_controlled_2015}).

Step-like features have been observed previously in works coupling BCS superconductors to inhomogeneous ferromagnetic textures. For example Witt \emph{et al.} in helical Ho/Nb bilayers \cite{witt_superconductivity_2012, chiodi_supra-oscillatory_2013}, Yi Zhu \emph{et al.} in GdN/Nb/GdN spin valves \cite{zhu2016superconducting}, and L.Y. Zhu \emph{et al.} in striped domain (Co/Pt)$_n$/Nb multilayers where it appears that the step shape can be modified by defining a current path parallel (no inhomogeneity -- no step) or perpendicular (inhomogeneity -- step) to the stripe domains \cite{zhu_altering_2008}.



While the exact origin of the step is unknown, it appears linked to the S--F proximity effect in all examples above. In this work we observe, in the upper panel of FIG.~\ref{dTc}, that the height of the step as a fraction of the transition is field history dependent. This step height change occurs at a different field than the largest changes in $M_r$ and $\Delta T_c$. While the change in step height does not appear to be intrinsically linked to the change in $T_c$, it is still clearly linked to the magnetic state of the Er layer. This further supports that the origin of the step is a fundamental feature of the S--F proximity effect, requiring theoretical description.


\section{Conclusions} 

In summary, the remanent state of Er, when proximity coupled to a Nb superconductor, can have a strong influence on $T_c$. The application of magnetic field is able to change the metastable magnetic state of the Er from a conical to fan state. This modification results in a fundamental change to the shape and temperature of the superconducting transition to zero resistivity. 

We hope the observation of this unconventional effect proves fruitful for refinement of S--F theory, particularly with the lack of current description of the stepped transition observed in this (and many similar) systems. 

A shift in $T_c$ of 140~mK is much larger than previously observed for singlet domain wall effects and is comparable to the largest observed by the generation of the LRTC in the context of the superconducting spin valve with 3$d$ ferromagnets. This system fulfils the requirements for cryogenic memory based upon the proposed architecture of Oh \emph{et al.} \cite{oh_superconductive_1997}, and we offer this materials system as a candidate for future super-spintronic device application.   


\begin{acknowledgments}
The authors would like to thank the UK EPSRC (grant numbers: EP/J010634/1, EP/J010650/1, EP/I031014/1 and EP/J01060X/1) for their financial support. NS~acknowledges JEOL Europe and ISIS neutron and muon source for PhD funding. 

The data associated with this paper are openly available from the University of Leeds data repository. https://doi.org/10.5518/142

\end{acknowledgments}

\bibliography{library}

\end{document}